\documentclass[11pt]{article}
\usepackage{moriond}
\usepackage{graphicx}
\usepackage{amsmath,amssymb,amscd,amsfonts,bm}

\newcommand{\ME}{M}

\newcommand{\LA}{\mathrm{A}}
\newcommand{\LB}{\mathrm{B}}

\newcommand{\LF}{\mathrm{F}}
\newcommand{\LI}{\mathrm{I}}

\newcommand{\La}{\mathrm{a}}
\newcommand{\Lb}{\mathrm{b}}
\newcommand{\Lc}{\mathrm{c}}

\def\ket#1{\big|{#1}\big\rangle}
\def\bra#1{\big\langle{#1}\big|}

\def\sket#1{\big|{#1}\big)}
\def\sbra#1{\big({#1}\big|}
\def\sbrax#1{\big({#1}}        

\newbox\charbox
\newbox\slabox
\def\s#1{{      
        \setbox\charbox=\hbox{$#1$}
        \setbox\slabox=\hbox{$/$}
        \dimen\charbox=\ht\slabox
        \advance\dimen\charbox by -\dp\slabox
        \advance\dimen\charbox by -\ht\charbox
        \advance\dimen\charbox by \dp\charbox
        \divide\dimen\charbox by 2
        \raise-\dimen\charbox\hbox to \wd\charbox{\hss/\hss}
        \llap{$#1$}
}}
\begin{document}
\vspace*{4cm}
\title{STRUCTURE OF PARTON SHOWERS INCLUDING QUANTUM INTERFERENCE}

\author{DAVISON E. SOPER}

\address{Institute of Theoretical Science, University of Oregon,\\
Eugene, Oregon 97403 USA\\
and Theory Group, CERN, CH1211 Geneva 23, Switzerland\\
\vskip 0.3 cm
{\rm ZOLTAN NAGY}\\
\vskip 0.1 cm
Theory Group, CERN, CH1211 Geneva 23, Switzerland
}

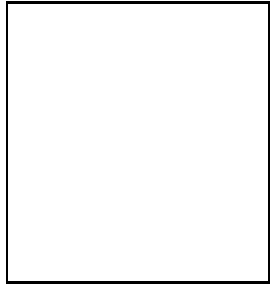
\maketitle\abstracts{
It is useful to describe a leading order parton shower as the solution of a linear equation that specifies how the state of the partons evolves. This description involves an essential approximation of a strong ordering of virtualities as the shower progresses from a hard interaction to softer interactions. If this is to be the only approximation, then the partons should carry color and spin and quantum interference graphs should be included. We explain how the evolution equation for this kind of a shower can be formulated. We discuss briefly our efforts to implement this evolution equation numerically.
}

\section{Introduction}

The evolution of a parton shower can be understood as a numerical solution of a linear evolution equation of the form\cite{NSshower}
\begin{equation}
\label{eq:calU}
\sket{\rho(t)} = {\cal U}(t,t_0)\sket{\rho(t_0)}
\end{equation}
with
\begin{equation}
\label{eq:evolution}
{\cal U}(t,t') = {\cal N}(t,t')
+ \int_{t'}^{t}\! d\tau\ 
{\cal U}(t,\tau)\,
{\cal H}_{\LI}(\tau)
\,{\cal N}(\tau,t')
\;\;.
\end{equation}
Here $\sket{\rho(t)}$ describes the system at ``shower time'' $t$, where increasing $t$ denotes increasingly soft interactions. The state at time $t$ is related to the state at an earlier time $t_0$ by a linear evolution operator ${\cal U}(t,t_0)$. The evolution equation~(\ref{eq:evolution}) is written using an operator ${\cal N}(t,t')$ that represents the probability for the system to go from time $t'$ to time $t$ with no splitting and another operator ${\cal H}_{\LI}(\tau)$ that represents a parton splitting to two partons. Thus the system either goes from $t_0$ to $t$ with no splitting or else it goes from $t_0$ to an intermediate time $\tau$ with no splitting, then splits at time $\tau$, then evolves from $\tau$ to $t$ with the full evolution operator, possibly involving further splittings.

In the simplest sort of shower, each parton carries a momentum $p$, so that $m$ partons carry momenta $\{p_1, p_2,\dots, p_m\} \equiv \{p\}_m$. We can denote the state in which there $m$ partons with momenta $\{p\}_m$ by $\sket{\{p\}_m}$. Then a general state $\sket{\rho}$ is a linear combination of basis states $\sket{\{p\}_m}$, with $\sbrax{\{p\}_m}\sket{\rho}$ representing the probability for there to be $m$ partons with momenta $\{p\}_m$. When the splitting operator ${\cal H}_{\LI}(\tau)$ acts on an $m$-parton state $\sket{\{p\}_m}$, it produces an $m+1$ parton state with a definite probability 
\begin{displaymath}
\sbra{\{\hat p\}_{m+1}}{\cal H}_{\LI}(\tau)\sket{\{p\}_m}
\;\;.
\end{displaymath}
There is a certain amount of freedom in specifying ${\cal H}_{\LI}(\tau)$, but one is constrained by the structure of the underlying quantum theory in the limit that the virtuality of the new pair of daughter partons approaches zero. In this simplest sort of shower, the basis states $\sket{\{p\}_m}$ are eigenstates of the no-splitting operator ${\cal N}(t,t')$,
\begin{equation}
{\cal N}(t,t')\sket{\{p\}_m}
= \Delta(t,t';\{p\}_m)\sket{\{p\}_m}
\;\;.
\end{equation}
The eigenvalue $\Delta(t,t';\{p\}_m)$ is the Sudakov factor
\begin{equation}
\Delta(t,t_0;\{p\}_{m}) =
\exp\left(-\int_{t_0}^{t} d\tau\ 
\frac{1}{(m+1)!}
\int\!\big[d\{\hat p\}_{m+1}\big]
\sbra{\{\hat p\}_{m+1}}{\cal H}_\LI(\tau)
\sket{\{p\}_{m}}\right)
\;\;.
\label{eq:sudakov0}
\end{equation}
The integrand
\begin{displaymath}
\frac{1}{(m+1)!}
\int\!\big[d\{\hat p\}_{m+1}\big]
\sbra{\{\hat p\}_{m+1}}{\cal H}_\LI(\tau)
\sket{\{p\}_{m}}
\end{displaymath}
is the total probability for the given state to split at time $\tau$. Then $\Delta(t,t_0;\{p\}_{m})$ is the probability for the state {\em not} to split between times $t_0$ and $t$.

\section{The parton shower in quantum chromodynamics}

So far, I have described a fairly general structure for a parton shower. This description might apply, for example, to \textsc{Pythia}.\cite{Pythia} The style of the description sketched above emphasizes that we are using a definite linear evolution equation, so that we can bring the full power of linear algebra to bear on the problem as needed. At this point, we need to expand the description so that it can encompasses quantum interference, spin, and color in quantum chromodynamics.\cite{NSshower}

To include quantum interference, the description should be based on the quantum amplitude. The quantum amplitude depends on the spins and colors of the partons, so we start with
\begin{displaymath}
\ME(\{p,f\}_{m})
^{c_\La,c_\Lb, c_1,\dots,c_m}_{s_\La, s_\Lb, s_1,\dots,s_m}
\;\;.
\end{displaymath}
Here, for hadron-hadron scattering, the partons carry labels $\La,\Lb, 1,\cdots,m$ and each parton has a momentum $p$, a flavor $f$, a spin index $s$ and a color index $c$. The array $\ME$ can be thought of as a vector in spin and color space,
\begin{displaymath}
\ket{\ME(\{p,f\}_{m})}
\;\;.
\end{displaymath}
The cross section for a possibly spin and color dependent observable $F$, including the proper factors for the parton distribution functions and for the number of color states $n_\Lc$ of each parton, is
\begin{equation}
\label{eq:opF}
\begin{split}
\sigma[F] = 
\sum_{m}&\frac{1}{m!}\int\big[d\{p,f\}_{m}\big]\,
\frac{f_{a/A}(\eta_{\La},\mu^2_\LF)\,
f_{b/B}(\eta_{\Lb},\mu^2_\LF)}
{4n_\Lc(a) n_\Lc(b)\,2\eta_{\La}\eta_{\Lb}p_\LA\!\cdot\!p_\LB}\,
\\&\!\times
\bra{\ME(\{p,f\}_{m})}
F(\{p,f\}_{m})
\ket{\ME(\{p,f\}_{m})}
\;\;.
\end{split}
\end{equation}
We rewrite $\sigma[F]$ in the form of a trace over the color\,$\otimes$\,spin space,
\begin{equation}
\label{eq:sigmaFtrace}
\begin{split}
\sigma[F] = 
\sum_{m}&\frac{1}{m!}\int\big[d\{p,f\}_{m}\big]\,
{\rm Tr}\{ \rho(\{p,f\}_{m})
F(\{p,f\}_{m})
\}
\;\;,
\end{split}
\end{equation}
where 
\begin{equation}
\label{eq:rhodef1}
\rho(\{p,f\}_{m}) = 
\ket{\ME(\{p,f\}_{m})}
\frac{f_{a/A}(\eta_{\La},\mu^{2}_{F})
f_{b/B}(\eta_{\Lb},\mu^{2}_{F})}
{4n_\Lc(a) n_\Lc(b)\,2\eta_{\La}\eta_{\Lb}p_\LA\!\cdot\!p_\LB}\,
\bra{\ME(\{p,f\}_{m})}
\;\;.
\end{equation}
Thus $\rho$ is the density operator in color\,$\otimes$\,spin space. We can expand $\rho(\{p,f\}_{m})$ in basis states $\ket{\{s,c\}_{m}}$ for the color\,$\otimes$\,spin space,
\begin{equation}
\label{eq:rhodef2}
\rho(\{p,f\}_{m})
= \sum_{s,c}\sum_{s',c'}
\ket{\{s,c\}_{m}}\,
\rho(\{p,f,s',c',s,c\}_{m})\,
\bra{\{s',c'\}_{m}}
\;\;.
\end{equation}
Thus $\rho(\{p,f,s',c',s,c\}_{m})$ is the density matrix. It is this matrix, with variable numbers of partons $m$, that is the basic object that evolves in a parton shower. In analogy with the notation used in the introduction, we consider $\rho$ to be a vector with
\begin{equation}
\label{eq:rhoket}
\rho(\{p,f,s',c',s,c\}_{m}) = \sbrax{\{p,f,s',c',s,c\}_{m}}\sket{\rho}
\;\;.
\end{equation}
Notice that each parton is described by its momentum, its flavor, two spin indices, and two color indices.

With this formulation, we can define\,\cite{NSshower} a splitting operator ${\cal H}_{\LI}(t)$ based on the behavior of the amplitude when two partons become collinear or one becomes soft. This gives a shower evolution equation of the form (\ref{eq:evolution}). However, in general the no-splitting operator ${\cal N}(t,t')$ is now a matrix in the color space.

\section{Issues of implementation} 
\label{sec:implementation}

A parton shower program like \textsc{Pythia}\cite{Pythia} starts with a state $\sket{\rho(t_0)}$ with just a few partons and produces states $\sket{\{\hat p,\hat f\}_m}$ with many partons at a final shower time $t_{\rm f}$. A parton shower program could also report a weight $w$ for the state. The weight times the probability to produce state $\{\hat p,\hat f\}_m$  is
\begin{equation}
\sbra{\{\hat p,\hat f\}_m}{\cal U}(t_{\rm f},t_0)\sket{\rho(t_0)}
\;\;.
\end{equation}
How can the evolution equation discussed above be implemented as a computer program? The evolution equation, solved iteratively, produces results expressed as integrals, so one could simply perform the integrations by numerical Monte Carlo integration, producing events and accompanying weights. However, for a large number of splittings it is likely that the fluctuations in the weights are too large for this most straightforward method to be practical. 

To proceed, we need to make some further approximations, with the understanding that any approximations should allow one to approach the exact solution of the evolution equation by using a sequence of approximations that become more and more exact as one proceeds, presumably at the cost of requiring more and more computer power. 

The base approximation is to average over spins and take the leading color approximation, $1/N_\Lc^2 \to 0$, where $N_\Lc = 3$ is the number of colors. With these approximations, we find\,\cite{NSspinless} that the evolution equation has the proper form to be implemented as a Markov process. In terms of numerical integration, this means that the integrals are nested and one can take the weights to be 1.

Next, we need to put spin back.\cite{NSspin} We assume that the final measurement function does not measure parton spin. However, the spin states of intermediate partons can influence the angular distributions of splittings.
If we use the spin-averaged shower to generate events, then the probability to generate a given shower history is wrong by the ratio of the splitting probabilities with spin to those without spin. We can take that ratio to be a weight that accompanies the event. Following an insight of Collins,\cite{JCCspin} we find that the spin weight factor can be calculated efficiently, using computational resources that are linear in the number of partons. (\textsc{Herwig} incorporates some spin effects using a related method.\cite{HerwigSpin})  Possibly, for reasons of numerical convergence, one should include the spin exactly for the first $N$ splittings, then average over spins for further splittings. Then the exact result is approximated more and more closely as we take $N$ to be large.

Finally, we need to put color back. This is more complicated than putting spin back. We expect to use a base approximation that is much less restrictive than the leading color approximation but that still allows efficient computation. There is a difference between the exact ${\cal H}_\LI$ and the approximate one. This difference would be included perturbatively at whatever order is needed to obtain an accurate result. Our work on color is in progress.

\section*{Acknowledgments}
This work was supported in part the United States Department of Energy and by the Hungarian Scientific Research Fund grant OTKA T-60432.


\section*{References}
\begin{thebibliography}{99}

\bibitem{NSshower}
  Z.~Nagy and D.~E.~Soper,
  JHEP {\bf 0709} (2007) 114
  [arXiv:0706.0017 [hep-ph]].
  
\bibitem{Pythia}
T.~Sjostrand, S.~Mrenna and P.~Skands,
  JHEP {\bf 0605} (2006) 026
  [arXiv:hep-ph/0603175];
  arXiv:0710.3820 [hep-ph].

\bibitem{NSspinless}
  Z.~Nagy and D.~E.~Soper,
  JHEP {\bf 0803} (2008) 030
  [arXiv:0801.1917 [hep-ph]].

\bibitem{NSspin}
  Z.~Nagy and D.~E.~Soper,
  arXiv:0805.0216 [hep-ph].
    
\bibitem{JCCspin}
  J.~C.~Collins,
  Nucl.\ Phys.\  B {\bf 304} (1988) 794.

\bibitem{HerwigSpin}
  P.~Richardson,
  JHEP {\bf 0111} (2001) 029
  [arXiv:hep-ph/0110108].

  

\end{thebibliography}
\end{document}